\documentclass[12pt,superscriptaddress,preprint,showpacs,nofootinbib,amsmath,amssymb,amsfonts]{revtex4-1}

\usepackage[T1]{fontenc}
\usepackage[latin9]{inputenc}
\usepackage{color}

\usepackage{ifpdf}
\ifpdf   %if using pdfLaTeX in PDF mode
    \usepackage[pdftex]{graphicx}
    \usepackage[hyperfootnotes=false,pdfstartview=,colorlinks=true,linkcolor=purple,citecolor=blue,urlcolor=blue,bookmarks=false]{hyperref}    \DeclareGraphicsExtensions{.png,.pdf}
    \usepackage{pgf}
\else    %if using LaTeX or pdfLaTeX in DVI mode
    \usepackage[usenames,dvipsnames]{color}
    \usepackage[hyperfootnotes=false,pdfstartview=,colorlinks=true,linkcolor=BrickRed,citecolor=blue,urlcolor=blue,bookmarks=false]{hyperref}
    \usepackage{graphicx}
    \DeclareGraphicsExtensions{.eps}   %\DeclareGraphicsExtensions{.eps,.bmp}
\fi

\hypersetup{pdftitle={Infationary Perturbations in Anisotropic, Shear-Free Universes},pdfauthor={Thiago S. Pereira, Saulo Carneiro, Guillermo A. Mena Marugan},pdfkeywords={Cosmology, Perturbation Theory, Inflation},pdfsubject={Theoretical Physics}}

\usepackage{color}

\newcommand{\be}{\begin{equation}}
\newcommand{\ee}{\end{equation}}
\newcommand{\bea}{\begin{eqnarray}}
\newcommand{\eea}{\end{eqnarray}}
\newcommand{\HH}{\mathcal{H}}
\newcommand{\nn}{\nonumber}

\begin{document}

\title{Inflationary Perturbations in Anisotropic, Shear-Free Universes}

\author{Thiago S. Pereira}
\email{tspereira@uel.br}
\affiliation{Departamento de F\'isica, Universidade Estadual de Londrina,
 86051-990, Londrina, Paran\'a, Brazil}
\affiliation{Institute of Theoretical Astrophysics, University of Oslo, M-0315 Oslo, Norway}

\author{Saulo Carneiro}
\affiliation{Instituto de F\'isica, Universidade Federal da Bahia, 40210-340, Salvador, Bahia, Brazil}
\affiliation{International Centre for Theoretical Physics, Strada Costiera 11, 34151, Trieste, Italy.\footnote{Associate Member}}

\author{Guillermo A. Mena Marugan}
\affiliation{Instituto de Estructura de la Materia, IEM-CSIC, Serrano 121, 28006 Madrid, Spain}

\date{\today}

%-------------------------------------------
% Abstract
%-------------------------------------------
\begin{abstract}
In this work, the linear and gauge-invariant theory of cosmological perturbations in a class of anisotropic and shear-free spacetimes is developed. After constructing an explicit set of complete eigenfunctions in terms of which perturbations can be expanded, we identify the effective degrees of freedom during a generic
slow-roll inflationary phase. These correspond to the anisotropic equivalent of the standard Mukhanov-Sasaki variables. The associated equations of motion present a remarkable resemblance to those found in perturbed Friedmann-Robertson-Walker spacetimes with curvature, apart from the spectrum of the Laplacian, which exhibits the characteristic frequencies of the underlying geometry. In particular, it is found that the perturbations cannot develop arbitrarily large super-Hubble modes.
\end{abstract}

\maketitle

%-------------------------------------------
% Introduction
%-------------------------------------------
\section{Introduction}
The unprecedented isotropy of the cosmic microwave background radiation revealed by modern CMB experiments is -- justifiably enough -- usually taken as an indication that we inhabit an isotropic universe. Together with the apparent (but far more debatable) homogeneity of matter distribution at scales
$\gtrsim 100h^{-1}\rm{Mpc}$, this observation suggests a generalization of the Copernican principle:
our universe is, on average, homogeneous and isotropic. At present, the Copernican principle remains a hallmark of cosmology, and its modification would profoundly alter the way we understand the physical universe. As a matter of fact, challenges to this scenario have largely boosted research in cosmology in the past, and will certainly do so in the foreseeable future.

In the last decade, new possibilities for a different approach to the cosmological principle have emerged from two different types of observations. On the one hand, the detection of the accelerated expansion of the universe \cite{Riess:1998cb,Perlmutter:1998np,Riess:2004nr} has motivated dark-energy-free cosmological models in which our (allegedly) privileged position in the universe would be perceived as a global cosmic acceleration \cite{Clarkson:2007pz,Uzan:2008qp}. On the other hand, discovered large-angles Cosmic Microwave Background (CMB) anomalies are being interpreted as a manifestation of a preferred direction in the universe (see Ref. \cite{Copi:2010na} for a recent review). Regardless of which description might turn out to be the best, these investigations are certainly well founded from the observational point of view -- not to mention that they constitute a necessary step to be faced in the so-called era of precision cosmology.

From a theoretical perspective, it is legitimate to ask when one can establish a one-to-one relation between our observations and the symmetries of the universe we live in. Although a homogeneous and isotropic universe would certainly lead to the observed CMB, is the converse true? Actually, Ehlers, Geren and Sachs (EGS) proved that the converse holds in an expanding universe provided that the isotropic CMB is measured by fundamental observers in a dust universe \cite{Ehlers:1966ad}. However, there are important counterexamples which allow the engineering of a richer class of models without jeopardizing well-established cosmological observations \cite{FMP,CE,Clarkson:1999yj}. The theorem demonstrated by EGS admits generalizations to a matter content given by a generic perfect fluid \cite{FMP}, and for an almost isotropic CMB \cite{SME}, case for which the associated geometry is only approximately homogeneous and isotropic. But, how can we phenomenologically conceive an anisotropic model of the universe in which basic observables, like for instance the redshift, remain isotropic, at least at dominant order? One example was proposed by Mimoso and Crawford in 1993 \cite{Mimoso:1993ym}. They showed that some anisotropic spacetimes, namely Bianchi type-III (BIII) and Kantowski-Sachs (KS) spacetimes, admit a shear-free expansion in which the scale factor has the same dynamics as a spatially curved Friedmann-Robertson-Walker (FRW) universe, provided that the anisotropic stress-tensor of the cosmological budget is in direct proportion to the electric part of the Weyl tensor. Since the dynamics is governed by a single scale factor, the metric can be brought to a conformally static form. It then follows by another EGS result that the radiation will be isotropic, in accordance with basic observations \cite{Ehlers:1966ad} (see also \cite{Clarkson:1999yj}). Further and independent developments have shown that this scenario can also be realized by endowing the universe with other forms of matter fields. In Ref. \cite{Koivisto:2010dr}, for example, a shear-free model with anisotropic curvature was constructed by means of a two-form field. Another interesting possibility was proposed in Ref. \cite{Carneiro:2001fz} (and further discussed in Ref. \cite{cmm}), where a massless and anisotropic scalar field was used to counterbalance the curvature of a BIII universe, producing an anisotropic model with a FRW-like background dynamics. This latter example is particularly interesting, given the role that scalar fields play in modeling early-universe physics.

Evidently, this alternate scenario may become appealing and competitive only if one is able to extract distinctive cosmological signatures from it. A possible observable at the level of the background geometry is the luminosity distance function: because the geometry traversed by photons along our line of sight is characterized by an anisotropic (spatial) curvature, the luminosity distance function will not have a direct correspondence to the standard one used in FRW models. One can then envisage a setting in which supernova surveys can be used to detect anisotropic modulations of the luminosity function
\cite{Koivisto:2010dr}, therefore rendering the proposed model falsifiable. However, if one wants to confront this theoretical scenario against cosmological data, then a much broader class of observables can be attained by considering perturbation theory, rather than focusing on the background. In particular, the whole framework of the inflationary era and its connection to structure formation provides a vast arena within which one can discuss the physical consequences of anisotropic models with isotropic expansion and put them to the test.

The theory of linear perturbations in non-FRW cosmological backgrounds has been an active field of research by its own, where many different models have been studied. The case of isotropic and inhomogeneous
Lema\^itre-Tolman-Bondi spacetimes, for example, has been recently worked out in Ref.
\cite{Clarkson:2009sc}, while the theory of perturbations in homogeneous and globally anisotropic Bianchi I spacetimes has been presented in Refs. \cite{Tomita:1985me,Pereira:2007yy,Gumrukcuoglu:2007bx}. Quite independently, the dynamics of linear perturbations in spatially curved anisotropic backgrounds has been investigated in Ref. \cite{Vasu:2008zz}, where the BIII case has been considered, and also in Ref. \cite{Zlosnik:2011iu}, where both the BIII and KS cases have been analyzed.

This paper presents the formalism necessary to extract the signatures that shear-free anisotropic models of the BIII family would generically produce in linear cosmological perturbations, specialized to situations where the universe is dominated by a slowly-varying scalar field, as it typically happens during inflation. With this aim, as well as to make more precise statements such as ``up to conditions on the behavior at infinity'', which are usually employed in a rather vague sense, we introduce an explicit set of complete eigenfunctions in terms of which perturbations can be expanded. This is a crucial step in perturbation theory that is sometimes overlooked in the literature, and unavoidable if we want to apply perturbation theory to specific cosmological models. To the best of our knowledge, these eigenfunctions have not been considered elsewhere.

As we will show, during a generic slow-roll inflationary phase, the dynamics of the perturbations can be described by effective degrees of freedom which have a remarkable resemblance to the standard Mukhanov-Sasaki (MS) variables \cite{Mukhanov:1985rz,Sasaki:1986hm}. However, since the eigenmodes of the Laplacian on the spatial sections are not given by standard plane-waves, we might expect non-trivial features in the morphology of both the temperature and matter fields of the universe. A particular signature which arises immediately from the spectrum of the Laplacian is the obstruction to have arbitrarily large super-Hubble modes. Because the curvature naturally introduces a spatial scale, the largest possible wave modes will be comparable to this scale. We also comment on the possibility of detecting specific features of shear-free anisotropy in the spectrum of gravitational waves.

This paper is structured as follows. We begin by reviewing the fundamental ideas behind anisotropic shear-free spacetimes in Sec. \ref{sfbd}, where we also introduce basic notations and definitions. In Sec.
\ref{lpt} we discuss the appropriate mode decomposition of the perturbations and use it to construct gauge-invariant variables. We close this section by determining a convenient basis of eigenfunctions of the Laplace operator. The dynamical equations of the cosmological perturbations are presented in Sec. \ref{me}, where we also construct the anisotropic equivalent of the MS variables. We conclude in Sec.
\ref{conclusion}. In this work we will adopt units such that $c=1=8\pi G$, and spacetime signature $(-,+,+,+)$. We follow the standard convention that Greek and Latin indices label spacetime and spatial coordinates, respectively. The spacetimes we will consider have a residual symmetry along one direction.
We choose this preferred direction to lie along the $z$-axis and let the indices $\{a,b,c\}$ run from 1 to 2, so as to span the $xy$-plane.

%-------------------------------------------
% Background Dynamics
%-------------------------------------------
\section{Shear-Free Background Dynamics}
\label{sfbd}

The simplest example of anisotropic universe is the one described by a Bianchi I spacetime. Since in this family of models the shear is entirely due to the anisotropic expansion of the universe, the spacetime cannot be both anisotropic and shear-free. This behavior is of course expected, since Bianchi I models contain the flat FRW case in the shear-free limit. The possibility of having anisotropic shear-free models with isotropic expansion relies on the fact that some anisotropic spacetimes do not possess a FRW limit. In these cases, the anisotropy of the space does not have to be encoded in the shear, but can instead be caused by imperfections of other kind in the fluid filling the universe.

An example of such spacetime is found in the BIII type. The BIII spacetime in question is spatially curved and possesses residual isotropy with respect to one direction. We choose the preferred axis of symmetry to point along the $z$ direction. Spatial sections are described by the direct product of the real line with a pseudo-sphere, $\mathbb{R}\times\mathbb{H}^2$, thus providing a space with negative constant curvature. In this work we will adopt a coordinate system in which the shear-free BIII metric can be written as:
\be
\label{background-metric}
{\rm d}s^2=a(\eta)^2\left(-{\rm d}\eta^2+\gamma_{ab}{\rm d}x^a{\rm d}x^b+{\rm d}z^2\right)\,,
\ee
where $a(\eta)$ is the isotropic scale factor, $\eta=\int{\rm d}t/a$ is the conformal time, and
$\gamma_{ab}$ is the standard metric of $\mathbb{H}^2$, which in a convenient coordinate system
$x^a=\{x,y\}$ reads:
\be
\label{b3}
\gamma_{ab}{\rm d}x^a{\rm d}x^b=dx^2+e^{2x}dy^2\,.
\ee
Following \cite{Carneiro:2001fz}, we will consider a matter sector composed by the standard inflaton
field $\varphi$ plus a massless scalar field $\vartheta$:
\bea
T^{\mu}_{(\varphi)\;\nu}&=&\partial^\mu\varphi\partial_\nu\varphi-\delta^\mu_{\;\nu}
\left(\frac{1}{2}\partial^\lambda\varphi\partial_\lambda\varphi+V(\varphi)\right)\,,\\
T^{\mu}_{(\vartheta)\;\nu}&=&\partial^\mu\vartheta\partial_\nu\vartheta
-\delta^\mu_{\;\nu}\left(\frac{1}{2}\partial^\lambda\vartheta\partial_\lambda\vartheta\right)\,.
\eea
The inflaton field being homogeneous, $\varphi=\varphi(\eta)$, satisfies the standard Klein-Gordon equation:
\be
\label{kgvarphi}
\varphi''+2\HH\varphi'+a^2V_{,\varphi}=0\,.
\ee
The prime denotes the derivative with respect to the conformal time. The massless field, on the other hand, obeys
\be
\label{kgphi}
\vartheta''+2\HH\vartheta'-\nabla^2\vartheta=0\,.
\ee
Here, $\nabla^2=D^aD_a+\partial^2_3$ and $D_a$ is the covariant derivative compatible with $\gamma_{ab}$. It is easy to check that the choice $\vartheta=Cz$, where $C$ is an arbitrary constant, not only solves Eq. (\ref{kgphi}), but also produces a diagonal  energy-momentum tensor:
\[
\rho_{(\vartheta)}=p_{(\vartheta)}^{\bot}=-p_{(\vartheta)}^{\parallel}=\frac{C^2}{2a^2}\,,
\]
where $p_{(\vartheta)}^{\bot}$ and $p_{(\vartheta)}^{\parallel}$ represent the pressures perpendicular
and parallel to the $xy$-plane, respectively. With these definitions, we can proceed and obtain
Einstein equations:
\bea
\label{G11}
3\HH^2&=&a^2\rho_{(\varphi)}+\frac{C^2}{2}+1\,, \\
\label{G22}
\HH^2+2\HH'&=&-a^2p_{(\varphi)}+\frac{C^2}{2}\,, \\
\label{G33}
\HH^2+2\HH'&=&-a^2p_{(\varphi)}-\frac{C^2}{2}+1\,.
\eea
It is then clear that if we choose $C^2=1$,\footnote{The ambiguity in the sign of $C$ comes from
the freedom in orienting the $z$-axis, which can be chosen at will.} Eqs. (\ref{G22}) and (\ref{G33}) turn out to coincide, while the whole system of equations becomes the standard Friedmann equations with curvature $\kappa=-1/2$:
\bea
\label{G00frw}
\HH^2&=&\frac{1}{3}\left(\frac{\varphi'^2}{2}+a^2V(\varphi)\right)-\kappa\,, \\
\label{Gijfrw}
\HH'&=&-\frac{1}{3}(\varphi'^2-a^2V(\varphi))\,.
\eea

This is indeed remarkable, since the dynamics is essentially that of an isotropic FRW universe with a perfect fluid source. On the other hand, it is worth noticing that our discussion surpasses the model obtained with our specific choice of solution for the scalar field; indeed, the same results can be obtained by endowing the universe with matter fields sustained by other considerations, such as a two-form field \cite{Koivisto:2010dr} or an imperfect fluid with stress \cite{Mimoso:1993ym}. In this work we
will adopt a rather pragmatic route and consider the solution $\vartheta=Cz$ as a way to unveil the generic signatures of inflationary perturbations in an anisotropic spacetime, while experiencing isotropic dynamics at the background level. As we will see, the main signatures of an inflationary era are rather insensitive to the perturbations of this field.

In order to find reduced dynamical variables for the perturbations, it will prove convenient to work
with Eqs. (\ref{G00frw}) and (\ref{Gijfrw}) combined in the following form:
\be
\label{H2Hprime}
\frac{\HH'}{\HH^2}=1+\frac{2\kappa-\varphi'^2}{2\HH^2}\,.
\ee
Finally, let us also introduce here the standard slow-roll parameters, which will be needed to identify
the effective degrees of freedom during inflation. They are defined by:

\be
\epsilon\equiv1-\frac{\HH'}{\HH^2}\,,\qquad \delta\equiv1-\frac{\varphi''}{\HH\varphi'}\,.
\ee
The first order slow-roll approximation is defined by the requirements that $\epsilon$ and
$\delta$ remain small during inflation \footnote{The constancy of $\epsilon$ and $\delta$ is automatically ensured because their time derivatives are of second order in these parameters \cite{Peter:1208401}.}:
\be
\label{sr}
\epsilon\ll 1\,, \qquad \delta\ll 1\,.
\ee
This is always true provided that the inflaton evolves under a considerably flat potential. In this regime, one can also show that:
\be
\label{eta-sr}
\HH\approx-\frac{1}{\eta}\frac{1}{1-\epsilon}\,,
\ee
where, we remind the reader, $\eta=\int{\rm d}t/a=\int{\rm d}a/(a\HH)$.

%-------------------------------------------
% Linear Perturbation Theory
%-------------------------------------------
\section{Linear Perturbation Theory}
\label{lpt}

We start by describing the general framework for the consideration of linear perturbations. The metric
and its inverse are written as a background piece plus contributions treated as perturbations:
\be
\label{gen-pert-metric}
g_{\mu\nu}\rightarrow g_{\mu\nu}+\delta g_{\mu\nu}\,,\qquad
g^{\mu\nu}\rightarrow g^{\mu\nu}+\delta g^{\mu\nu}\,.
\ee
At linear order in perturbation theory, this implies that
\[
\delta g^{\mu\nu}=-g^{\mu\alpha}g^{\nu\beta}\delta g_{\alpha\beta}\,.
\]
To manipulate indices in perturbed quantities, one should apply (\ref{gen-pert-metric}) and then discard second order terms in the perturbations.

The matter sector is composed by the fields $\vartheta$, $\varphi$, and their perturbations, which can
be written as:
\bea
\vartheta&\rightarrow&\vartheta+\delta\vartheta\,,\\
\varphi&\rightarrow&\varphi+\delta\varphi\,.
\eea

In the following subsections we will discuss the general parametrization of $\delta g_{\mu\nu}$ and the construction of gauge-invariant variables.

\subsection{Mode decomposition}

The general construction of perturbation theory in cosmology is usually based on the identification of a set of independent modes representing the degrees of freedom of the perturbations. For spacetimes admitting a 1+3 foliation, these modes are typically identified with irreducible pieces of the symmetry group of the corresponding 3-dimensional spatial sections. In the FRW case, these are maximally symmetric spaces, and a unique decomposition of symmetric rank-2 tensors in terms of Scalars, (transverse) Vectors and (transverse-traceless) Tensors is always possible (SVT decomposition, from now on) \cite{Stewart:1990fm}. In the cases studied here, the constant-time hypersurfaces are not maximally symmetric spaces, so the standard SVT decomposition is not adequate. However, since the considered BIII spacetimes have a 2-dimensional maximally symmetric subspace, namely the pseudo-sphere, we can instead proceed with a 1+2+1 decomposition of the dynamical modes.

Some important differences with the standard SVT decomposition are immediate: given that the largest subspace is of dimension 2, independent transverse and traceless symmetric tensor modes cannot exist on it, and the most general symmetric rank-2 tensor is decomposed as
\be
\label{tensor-SV}
T_{ab}=X\gamma_{ab}+D_{a}D_{b}Y+D_{(a}\bar{Z}_{b)}\,,\qquad D^a\bar{Z}_{a}=0\,,
\ee
where $X$ and $Y$ are scalars, $\bar{Z}_a$ is a transverse vector, and we recall that $D_a$ is the covariant derivative in $\mathbb{H}^2$. Likewise, any vector $V_a$ will be decomposed as
\be
\label{vector-SV}
V_{a}=D_aV+\bar{V}_a\,,\qquad D^a\bar{V}_{a}=0\,,
\ee
where $V$ is a scalar and $\bar{V}_a$ is again transverse. Notice that the transversality condition eliminates one of the two variables of the transverse vectors, so that they are described by one single scalar function. Nonetheless we will maintain the standard terminology and refer to the splitting
(\ref{tensor-SV}) and (\ref{vector-SV}) as a Scalar-Vector (SV) decomposition. In the following, every barred vector will refer to a transverse vector.

Based on the discussion above, we can parametrize the most general perturbation $\delta g_{\mu\nu}$ as
\be
\label{perturbed-metric}
{\rm d}s^2=a^2\left[-(1+2A){\rm d}\eta^2+2B_a{\rm d}x^a{\rm d}\eta+2E_a{\rm d}x^a{\rm d}z
+2C{\rm d}\eta{\rm d}z+h_{ab}{\rm d}x^a{\rm d}x^b+2F{\rm d}z^2\right]\,,
\ee
where
\bea
B_{a}&=&D_aB+\bar{B}_a\,,\\
E_{a}&=&D_aE+\bar{E}_a\,,\\
h_{ab}&=&2S\gamma_{ab}+2D_aD_bU+2D_{(a}\bar{V}_{b)}\,.
\eea
This decomposition is unique, and the introduced functions and vectors completely specify the 10 degrees
of freedom in $\delta g_{\mu\nu}$.

\subsection{Gauge-Invariant Variables}

Let us specify the gauge transformations of the above free functions. Under
an infinitesimal gauge transformation of the form
\be
\label{gauge}
\tilde{x}^\mu=x^\mu-\xi^\mu\,,
\ee
any perturbed quantity $\delta Q$ transforms as
\be
\label{gauge-transf}
\delta Q\rightarrow \delta\tilde{Q}=\delta Q + \mathcal{L}_\xi \underline{Q}\,,
\ee
where $\underline{Q}$ denotes the background quantity to be perturbed. By definition, gauge-invariant variables are those for which
\be
\mathcal{L}_\xi \underline{Q}=0\,.
\ee
According to the 1+2+1 splitting of the spacetime, the gauge vector $\xi^\mu$ will be parametrized by
\be
\label{xi}
\quad\xi^\mu=(T,D^aL+\bar{L}^a,J)\,,
\ee
where $T$, $L$, and $J$ are scalars and $\bar{L}_a$ is a transverse vector.

\subsubsection{Metric}

If the quantity $\underline{Q}$ is the spacetime metric, then the gauge transformation
(\ref{gauge-transf}) becomes
\[
\mathcal{L}_\xi g_{\mu\nu}=2\nabla_{(\mu}\xi_{\nu)}\,,
\]
where $\nabla$ is the full covariant derivative compatible with the background metric
(\ref{background-metric}) and, as usual, the parentheses denote symmetrization of indices. Using Eqs.
(\ref{perturbed-metric}) and (\ref{xi}), we find the following gauge transformations for the scalar variables:
\bea
A &\rightarrow& A+T'+\HH T\,,\\
B &\rightarrow& B-T+L'\,,\\
C &\rightarrow& C-\partial_z T+J'\,,\\
S &\rightarrow& S+\HH T\,,\\
U &\rightarrow& U+L\,,\\
E &\rightarrow& E + \partial_z L + J\,,\\
F &\rightarrow& F+\HH T + \partial_z J\,.
\eea
In a similar manner, vector modes are found to transform as:
\bea
\bar{B} &\rightarrow& \bar{B}+\bar{L}'\,,\\
\bar{V} &\rightarrow& \bar{V}+\bar{L}\,,\\
\bar{E} &\rightarrow& \bar{E}+\partial_z\bar{L}\,.
\eea

In total we have seven scalar modes $(A,B,C,S,U,E,F)$ and three transverse vector modes
$(\bar{B},\bar{E},\bar{V})$, for a total of ten free functions in $\delta g_{\mu\nu}$. We can
now use the four free functions in $\xi^\mu$ to construct six gauge-invariant variables.
One possibility is
\bea
\label{PhiGI}
\Phi &\equiv& A+\frac{1}{a}[a(B-U')]'\,,\\
\Psi &\equiv& S+\HH (B-U')\,,\\
\Pi &\equiv& -C + E'+ \partial_z(B-2U')\,,\\
\label{LambdaGI}
\Lambda &\equiv& F+\HH(B-U')-\partial_z(E-\partial_z U)\,,
\eea
for the scalar sector, and
\bea
\label{GammaGI}
\bar{\Gamma}_a &\equiv& \bar{V}_a'-\bar{B}_a\,,\\
\label{OmegaGI}
\bar{\Omega}_a &\equiv& \bar{E}_a-\partial_z\bar{V}_a\,,
\eea
for the vector sector. Other variables exist, like for example $\bar{\Sigma}_a\equiv\bar{E}_a'-\partial_z\bar{B}_a$. But they are not functionally independent; for instance, it is easy to check that
$\bar{\Sigma}_a=\bar{\Omega}_a'+\partial_z\bar{\Gamma}_a$.

\subsubsection{Matter}

We will consider the perturbations to both the inflaton and the anisotropic scalar field. Under the gauge transformation (\ref{xi}), any perturbation to $\vartheta$ and $\varphi$ will transform respectively as
\bea
\mathcal{L}_\xi\vartheta&=&J\partial_z\vartheta\,,\\
\mathcal{L}_\xi\varphi&=&\varphi'T\,.
\eea
Combining these formulas with the metric potentials, we can define the following gauge-invariant
perturbations:
\bea
\label{gi-deltatheta}
\widetilde{\delta\vartheta}=\delta\vartheta-E+\partial_zU\,,\\
\label{gi-deltaphi}
\widetilde{\delta\varphi}=\delta\varphi+\varphi'(B-U')\,.
\eea
\\

Altogether, the system of variables composed by (\ref{PhiGI}-\ref{OmegaGI}) and
(\ref{gi-deltatheta}-\ref{gi-deltaphi}) span the space of cosmological perturbations, implying that
fully gauge-invariant linear equations are possible. However, in practice it is much easier to perform computations in some specific gauge and convert the final result to gauge-invariant expressions. For this purpose, a suitable choice is
\[
B=E=U=0=\bar{V}_a\,.
\]
This choice fixes the gauge completely and determines the free functions of $\delta g_{\mu\nu}$ and $\delta T_{\mu\nu}$ in terms of gauge-invariant variables. We can therefore perform our calculations in this gauge, and translate the final result into gauge-invariant quantities. This is always possible in linear theory due to the Stewart-Walker lemma \cite{Stewart:1974uz}.

\subsection{Eigenfunctions}

The natural home to define functions on the spatial sections, and in particular scalar perturbations, is the Hilbert space of square integrable functions with measure provided by the volume element
$e^{x}{\rm d}x{\rm d}y{\rm d}z$. Since the Laplacian is a self-adjoint operator on this space, we can obtain a spectral decomposition of the identity associated with it. This decomposition can then be employed to express any perturbation as a series in terms of eigenmodes of the Laplacian.

In the coordinates defined by Eq. (\ref{b3}) the Laplacian takes the following form:
\be
\label{three-laplacian}
\nabla^{2}=\partial_{x}+\partial_{x}^{2}+e^{-2x}\partial_{y}^{2}+\partial_{z}^{2}\,.
\ee
We now want to solve the eigenvalue problem
\[
\nabla^{2}\phi(x,y,z)=-q^{2}\phi(x,y,z)\,,\quad q\in\mathbb{R}\,.
\]
We recall that the solutions must be (generalized) states in our Hilbert space of square integrable functions. The problem can be solved by means of a simple separation of variables,
\[
\phi(x,y,z)=f(x)g(y)h(z)\,,
\]
in terms of which the eigenvalue problem becomes:
\[
\frac{1}{f}\left(\frac{df}{dx}+\frac{d^{2}f}{dx^{2}}\right)+e^{-2x}\frac{1}{g}\frac{d^{2}g}{dy^{2}}
+\frac{1}{h}\frac{d^{2}h}{dz^{2}}=-q^{2}\,.
\]
This equation has a nontrivial solution provided that
\be
\frac{1}{g}\frac{d^{2}g}{dy^{2}} = -m^{2}\,,\qquad
\frac{1}{h}\frac{d^{2}h}{dz^{2}} = -n^{2}\,,\qquad
\frac{d^{2}f}{du^{2}}+\left(\frac{q^{2}-n^{2}}{u^{2}}-1\right)f=0\,,
\ee
with real numbers $m$ and $n$ and $u\equiv me^{-x}$. The first two equations admit standard plane-wave solutions, while the third has a solution in terms of modified Bessel functions:
\be
f(u)=C_{1}\sqrt{u}I_{\nu}(u)+C_{2}\sqrt{u}K_{\nu}(u)\,,
\ee
with the eigenvalue $\nu$ given by
\be
\nu=\sqrt{\frac{1-4(q^{2}-n^{2})}{4}}\,.
\ee
For real values of $\nu$, the solutions $\sqrt{u}I_{\nu}(u)$ and $\sqrt{u}K_{\nu}(u)$ diverge at
$u\rightarrow\infty$ and $u\rightarrow 0$, respectively. Actually, these solutions are not appropriate for the description of linear perturbations, since one can check that they do not belong to the Hilbert space of functions under consideration \cite{bessel} (square integrable with the measure $e^x {\rm d}x$ or, equivalently, $du/u^2$), not even as generalized delta-normalizable states. However, we can find well-behaved (and real) solutions if we restrict $\nu$ to imaginary values. In other words, admissible
solutions are possible if
\be
\label{q-restriction}
q^2\geq\frac{1}{4}+n^2\,.
\ee

\begin{figure}
\begin{centering}
\includegraphics[scale=0.9]{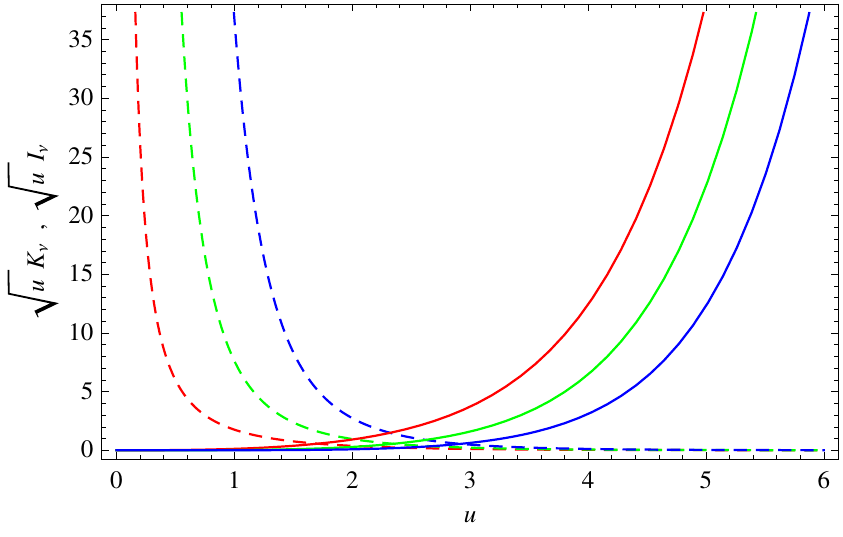}
\includegraphics[scale=0.9]{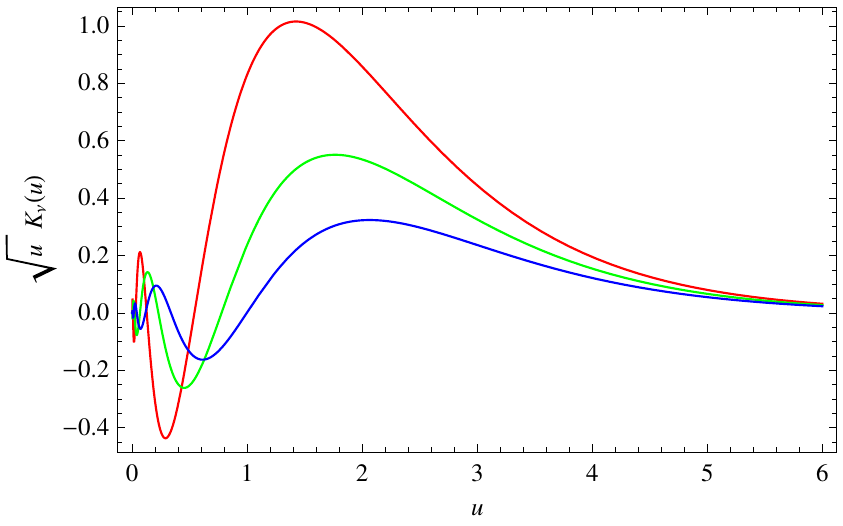}
\par
\end{centering}
\caption{Eigenfunctions of the Laplacian in Bianchi III spacetimes. The left panel shows
$\sqrt{u}I_\nu$ (continuous line) and $\sqrt{u}K_\nu$ (dashed line) for real values of $\nu$. The behavior of $\sqrt{u}K_\nu$ for imaginary $\nu$ is shown in the right panel. The values of $(q,n)$ used were: (0,2), (4,5), and (7,8) (left); (3,2), (4,3), and (5,4) (right).}
\label{fig1}
\end{figure}

We show some plots of these solutions for some representative values of $\nu$ in Fig. (\ref{fig1}). Note that the restriction (\ref{q-restriction}) implies a lower bound on the wavenumber $q$, which is equivalent to an upper bound on the associated wavelength $\lambda=2\pi/q$. If we consider the limiting case
$n=0$, we see that $q\geq1/2=|\kappa|$. In other words, at a given instant $\eta$, no physical wavelength can be bigger than $4\pi a(\eta)$. This is an important signature of the model which, in principle,
might be detected in CMB multipoles via the Grishchuk-Zel'dovich effect
\cite{Grishchuck:1978,GarciaBellido:1995wz}. We will however postpone this analysis to a forthcoming paper \cite{CMP:2012}.

Returning to the analysis of the eigenfunctions, we have now a complete set in terms of which
perturbations can be expanded:
\be
\phi_{(\omega,m,n)}(x,y,z)=\frac{\sqrt{2\omega\sinh(\pi\omega)}}{2\pi^2}e^{-x/2}
K_{\nu}(|m|e^{-x})e^{imy}e^{inz}\,,
\ee
where $\nu\equiv i\omega$ and where the constant prefactor ensures that these functions are delta-
normalized \cite{CMP:2012}. In conclusion, any perturbation $f(x,y,z)$ can now be expanded as:
\be
\label{eigenfunc}
f(x,y,z)=\int_0^\infty{\rm d}\omega\int_{-\infty}^\infty{\rm d}n\int_{-\infty}^\infty{\rm d}m
\;\phi_{(\omega,m,n)}(x,y,z)\widetilde{f}(\omega,m,n)\,,
\ee
its inverse being given by:
\be
\widetilde{f}(\omega,m,n)=\int_{\mathbb{R}^3}e^{x}{\rm d}x{\rm d}y{\rm d}z\;\phi^{*}_{(\omega,m,n)}(x,y,z)
f(x,y,z)\,.
\ee
The symbol $*$ denotes complex conjugation. Note also 
that, since $K_{\nu}(u)=K_{\nu^*}(u)$, only positive values of $\omega$ contribute to Eq.
(\ref{eigenfunc}).

%-------------------------------------------
% Master Equations
%-------------------------------------------
\section{Master Equations}
\label{me}

We are now in an adequate position to identify the effective degrees of freedom and write down the master equations for the perturbations. A generic feature of perturbation theory in anisotropic spacetimes containing shear is a coupling, already at first order in perturbations, between different perturbative modes \cite{Pereira:2007yy,Pitrou:2008gk}. Since we are working in a shear-free scenario, such couplings will not appear. However, we will still find a residual coupling between different scalar modes. In our case, this is due to the way scalar variables are divided in the 1+2+1 splitting, and also to the spatial curvature of the background spacetime.

The expressions for the fully linearized Einstein and energy-momentum tensors are presented in the Appendix. Here we explain how to gather their combinations into the final master equations.

\subsection{Scalar modes}
All linearized Einstein equations contain scalar modes which can be combined to derive master dynamical equations. Before we proceed, note that Eq. (\ref{dgab})  (with $a\neq b$) implies for the  scalar modes that
\be
\Lambda=-\Phi\,.
\ee
This extra constraint is similar to what happens in the FRW case when the matter source does not contain stress at first perturbative order (which is also the case considered here). This relation is always true provided that the spatial average of the perturbations vanishes (see Ref.
\cite{Mukhanov:1990me}). From now on, we will substitute this constraint in all the equations.

The first dynamical scalar equation can be obtained from the perturbed Klein-Gordon equation. Dropping
the tilde in definition (\ref{gi-deltaphi}) for convenience, we find from $\nabla_\mu T^{\mu}_{\;0}=0$ that
\be
\delta\varphi''+2\HH\delta\varphi'-\nabla^2\delta\varphi+a^2V_{,\varphi\varphi}\delta\varphi=
\varphi'(2\Phi'+4\HH\Phi-2\Psi'-\partial_z\Pi)+2\varphi''\Phi\,.	
\ee
Using the constraint (\ref{dg0a}) to replace $\partial_z\Pi$ in the expression above, this equation
reduces to:
\be
\label{kg}
\delta\varphi''+2\HH\delta\varphi'-(\nabla^2-a^2V_{,\varphi\varphi})\delta\varphi
=2\varphi'^2\delta\varphi+2\varphi''\Phi\,.	
\ee

The second scalar equation can be found by subtracting (\ref{dg00}) from  the trace of (\ref{dgab}). We then use the constraint (\ref{dg0a}) to substitute $\partial_z\Pi$ in terms of $\Phi$, $\Psi$, and
$\delta\varphi$. The resulting equation is:
\be
\label{X}
X''+2\HH X'-(\nabla^2+4\kappa)X=4\HH'\Phi-2\varphi''\delta\varphi\,.
\ee
Here $X\equiv\Psi-\Phi$. Eqs. (\ref{kg}) and (\ref{X}) are clearly coupled, and in principle one could expect that they were reducible to a single but closed dynamical equation involving some combination of
$\Psi$, $\Phi$, and $\delta\varphi$. However, we will show in Subsec. \ref{redeqs} that a closed equation for these variables is not generally attainable. Strictly speaking, we can reach an equation with those properties only during a slow-roll inflationary phase.

The dynamics of the scalar sector is also described by the equation for the variable $\Pi$. By
combining Eqs. (\ref{dg00}), (\ref{dg03}), (\ref{dg3a}), and the corresponding equations for 
$\delta T^\mu_{\;\nu}$ we find that:
\be
\label{Pi}
\Pi''+2\HH\Pi'-(\nabla^2-2\HH'+4\kappa)\Pi=0\,.
\ee
Finally, we have also the dynamics of the perturbed field $\delta\vartheta$. By combining its Klein-Gordon equation with Eq. (\ref{dg3a}) one gets (also dropping the tilde in the definition
(\ref{gi-deltatheta})):
\be
\label{theta}
\delta\vartheta''+2\HH\vartheta'-\nabla^2\vartheta-4\kappa\vartheta=0\,.
\ee
It is interesting to note that, even though the scalar sector is being sourced by the two matter fields, the above equation is completely decoupled from other perturbed variables and, in particular, from the perturbations of the inflaton. This property is reminiscent from the fact that $\vartheta$ is a massless field without dynamics at the background level.

\subsection{Vector modes}
The dynamics of vector modes are contained in the coupled system of equations (\ref{dgab}) and
(\ref{dg3a}). Because the constraint (\ref{dg0a}) does not allow us to decouple them, in principle we
have to treat $\bar{\Omega}_a$ and $\bar{\Gamma}_a$ implicitly as one degree of freedom. However, from Eq. (\ref{dgab})
we see that:
\be
\label{GammaOmega}
\partial_z\bar{\Omega}_a+\bar{\Gamma}'_a+2\HH\bar{\Gamma}_a=f(\eta,z)\zeta_a\,,
\ee
where $\zeta_a$ is a Killing vector of $\mathbb{H}^2$ and $f$ is arbitrary in principle. On the other hand, it is not difficult to check that the fact that the perturbations must belong to the space of square integrable functions with respect to the volume element defined by the spatial background metric implies that the vector $\zeta_a$ must be normalizable with respect to the measure $e^x {\rm d}x{\rm d}y$. But there exists no Killing vector on $\mathbb{H}^2$ with this property (see Eqs. (1.4) in Ref. \cite{Reboucas:1982hn}). Therefore, the only possibility left is that $f=0$. In this case we can use Eq. (\ref{GammaOmega}) to decouple the system of equations (\ref{dgab}) and (\ref{dg3a}), leaving us with one single equation for the variable
$\bar{\Omega}_a$:
\be
\label{Omega}
\bar{\Omega}_a''+2\HH\bar{\Omega}_a'-(\nabla^2+2\kappa)\bar{\Omega}_a=0\,.
\ee
Remarkably, this equation is formally the same as the equation for the polarizations of gravitational waves in a curved FRW universe, where the contribution from the curvature appears with opposite sign
(see Ref. \cite{Peter:1208401}). This difference in sign is due to the 1+2+1 splitting of the perturbative modes, as can be checked directly from the perturbed Einstein equations (see the Appendix).

\subsection{Reduced equations}\label{redeqs}
The system of equations (\ref{kg}-\ref{theta}) and (\ref{Omega}) fully characterizes the dynamics. Note that Eqs. (\ref{Pi}), (\ref{theta}), and (\ref{Omega}) are decoupled, and therefore represent three independent degrees of freedom of the system. Given that we have eight gauge-invariant variables (6 metric potentials plus two scalar field perturbations) and four constraints from Einstein equations, we would in principle expect that the coupled scalar equations (\ref{kg}) and (\ref{X}) could be combined into one single scalar equation. When inflation is driven by a single scalar field in a flat FRW background, this reduction is always possible and leads to the dynamical equation for the scalar MS variable. In the present case this reduction does not seem possible in its full generality. Curiously, this difficulty does not arise owing to the anisotropy of the spacetime, which generically couples perturbative modes, but rather by the presence of the spatial curvature. As a matter of fact, precisely the same difficulty appears in curved FRW universes, where a nonzero spatial curvature seriously compromises the direct identification of the scalar MS variable (see Ref. \cite{Garriga:1999vw}, though). Fortunately, the terms which prevent Eqs. (\ref{kg}) and
(\ref{X}) to be combined into one single dynamical equation are precisely the ones proportional to the slow-roll parameters. If inflation lasts for at least 60 e-folds, so as to solve the standard big bang problems, the dynamics of the scalar perturbations will be given by just one scalar MS-like variable. To see that, we introduce a new variable $v$ defined as
\be
v\equiv a\left(\delta\varphi-\frac{\varphi'}{2\HH}X\right)\,.
\ee
If we now combine Eqs. (\ref{kg}) and (\ref{X}), using the background equation (\ref{H2Hprime}) and the constraint (\ref{dg0a}), we get the following equation after a lengthy but straightforward calculation:
\be
v''-\nabla^2v-\left\{\frac{\bar{z}''}{\bar{z}}
+2\kappa\left[\frac{1}{\varphi'}\left(\frac{\varphi'}{\HH}\right)'+2\right]\right\}v=
\frac{a}{2}\left(\frac{\varphi'}{\HH}\right)'\left[\partial_z\Pi-\frac{X}{\HH}\right]\,,
\ee
where $\bar{z}=a\varphi'/\HH$.

We now employ the slow-roll coefficients, in terms of which we have
\bea
\frac{\bar{z}''}{\bar{z}}&=&\HH^2\left[2-3\delta+2\epsilon+\delta^2-\epsilon\delta
+\frac{\epsilon'-\delta'}{\HH}\right], \\
\left(\frac{\varphi'}{\HH}\right)'&=&\sqrt{2\HH^2\epsilon+\kappa}\;(\epsilon-\delta)\,.
\eea
These expressions are exact. Their expansion at first-order in a slow-roll approximation can be directly obtained by means of Eqs. (\ref{sr}) and (\ref{eta-sr}):
\bea
\frac{\bar{z}''}{\bar{z}}&\approx&\frac{1}{\eta^2}(2+6\epsilon-3\delta), \\
\left(\frac{\varphi'}{\HH}\right)'&\approx&\frac{\sqrt{2\epsilon}}{\eta}\;(\epsilon-\delta)\,.
\eea
where in the last equality we have assumed that $\kappa/\HH^2$ can be neglected compared to $\epsilon$, since during inflation $\HH=aH$ grows almost exponentially, whereas $\kappa$ remains constant. Taking the above expansions into consideration, we finally obtain:
\be
\label{eqv}
v''-\left[\nabla^2+\frac{1}{\eta^2}(2+6\epsilon-3\delta)\right]v=0\,.
\ee
This equation is remarkably close to the one found in the isotropic FRW case except, obviously, from the different spectrum of the Laplacian.

The remaining MS variables are much easier to obtain. With a suitable scaling of $\Pi$, $\vartheta$ and
$\Omega_a$, Eqs. (\ref{Pi}), (\ref{theta}), and (\ref{Omega}) can be rewritten as:
\bea
\label{equ-general}
u''&-&\left(\nabla^2-\frac{a''}{a}+2\HH^2+4\kappa\right)u=0\,,\\
\theta''&-&\left(\nabla^2-\frac{a''}{a}+4\kappa\right)\theta=0\,,\\
y_a''&-&\left(\nabla^2+\frac{a''}{a}+2\kappa\right)y_a=0\,,
\eea
where the new variables that have been introduced are explicitly given by:
\be
u\equiv a\Pi\,,\qquad\theta=a\delta\vartheta\,,\qquad y_a\equiv a\bar{\Omega}_a\,.
\ee
Differently from Eq. (\ref{eqv}), these equations are valid regardless of the slow-roll approximation.
For completeness, however, we present their expressions during this stage. Using:
\be
\frac{a''}{a}\approx\frac{2+3\epsilon}{\eta^2}\,,\qquad\HH^2\approx\frac{1}{\eta^2}(1+2\epsilon)\,,
\ee
and again neglecting $\kappa$ compared to ${\cal H}^2\epsilon\approx\epsilon/\eta^2$, the above equations become:
\bea
\label{equ}
u''&-&\left(\nabla^2+\frac{\epsilon}{\eta^2}\right)u=0\,,\\
\label{eqtheta}
\theta''&-&\left(\nabla^2-\frac{2+3\epsilon}{\eta^2}\right)\theta=0\,,\\
\label{eqy}
y_a''&-&\left(\nabla^2+\frac{2+3
\epsilon}{\eta^2}\right)y_a=0\,.
\eea
We remind the reader that, despite its appearance, $y_a$ is a transverse vector represented by one single (scalar) function.
\\

Equations (\ref{eqv}), and (\ref{equ}-\ref{eqy}) are the central result of this work, and govern the dynamics of the effective degrees of freedom of inflationary perturbations in the considered BIII background spacetime. Given that Eq. (\ref{eqv}) clearly describes the scalar perturbative mode, and that $u$ and $y_a$
are pure metric potentials, we can interpret Eqs. (\ref{equ}) and (\ref{eqy}) as representing the two polarizations of the gravitational waves. We notice that these two polarizations have different
dynamics, a fact which is a generic feature of anisotropic spacetimes.

As a final remark, we want to make some comments regarding the massless scalar field $\vartheta$ used
in this work. While its presence is necessary to ensure the FRW dynamics of the anisotropic background, its perturbation is completely decoupled from the perturbation of the inflaton field. In other words, equation (\ref{eqv}), which is the central equation regarding inflationary predictions, remains unchanged regardless of whether
we perturb or not  $\vartheta$. Nonetheless, had we chosen not to perturb this field,
the equation  for $u$ would not be given by (\ref{equ-general}), but rather by:
\be
u''-\left(\nabla^2-\frac{a''}{a}+2\HH^2\right)u=0\,,
\ee
as the reader can directly check using the equations in the Appendix and taking 
$\delta T^\mu_{(\vartheta)\;\nu}=0$. This indirect influence of $\delta\vartheta$ in 
$u$, as well as the different dynamics of the two polarizations of the gravitational waves, might, in principle, lead to observational signatures in their spectrum.

\section{Conclusion}
\label{conclusion}

The possibility that our current cosmological observations can be made compatible with a universe with anisotropic geometry is challenging but really exciting, since this would naturally lead us to consider extensions of the cosmological principle beyond our present basis for the standard cosmological model. In this context, a especially nice scenario is offered by anisotropic spacetimes which are conformally flat (shear-free), since they admit an isotropic background expansion, and therefore an isotropic redshift. In this work, we have unveiled the dynamics that cosmological perturbations would have in a BIII spacetime of this kind during a generic slow-roll inflationary phase. Based on a 1+2+1 splitting of the spacetime, we have constructed the full set of linear and gauge-invariant perturbations, from which a set of four canonical degrees of freedom have been identified, and described by variables of the MS type in the slow-roll regime. Interestingly, these canonical variables have a remarkable similarity to their isotropic (FRW) counterparts. It is worth pointing out that, although the identification of the scalar MS variable is only possible during a strict slow-roll inflationary phase, the same happens in spatially curved FRW universes. In this latter case, the identification of the scalar MS variable becomes severely obstructed by the appearance of non-local terms owing to the spatial curvature \cite{Garriga:1999vw}. We have shown here that these terms are of higher order in a slow-roll expansion, and can therefore be ignored during inflation.

In order to expand our equations in natural modes of our system, simplifying in this way their resolution and avoiding complicated couplings, we have constructed a complete set of eigenfunctions adapted to the symmetries of the spacetime. This also selects a specific functional space that provides the natural home for the perturbations, and allows us to deal with issues related with boundary conditions on these perturbations in a straightforward manner. A footprint of cosmological perturbations in this cosmological scenario is their inability to develop arbitrary super-Hubble modes. This seems to be a general feature of spaces with hyperbolic spatial sections, since the same happens in open FRW models as well. In this FRW case, analytic extensions of the eigenfunctions are usually invoked in order to access the physics of superhorizon perturbations \cite{GarciaBellido:1995wz}. Whether superhorizon modes will have an appreciable impact in future CMB observations or not, and whether such extensions are necessary in the present model is a question for further investigation.

%-------------------------------------------
% Acknowledgements
%-------------------------------------------
\acknowledgements
We would like to thank David Mota, Cyril Pitrou, Miguel Quartin, Mikjel Thorsrud and Jean-Philippe Uzan
for useful conversations during the realization of this paper. This work was supported by the Research Council of Norway under project No. 211124, by Conselho Nacional de Desenvolvimento Cient\'ifico e
Tecnol\'ogico (Brazil) under project No. 308758/2011-0, and by the research grants No. MICINN/MINECO FIS2011-30145-C03-02 and CPAN CSD2007-00042 from Spain. T.S.P. thanks the Instituto de Estructura de la Materia of the CSIC in Spain and the Institute for Theoretical Astrophysics of Norway for their hospitality during the development of this work.

\newpage
%-------------------------------------------
% Appendix
%-------------------------------------------
\appendix

\section{Useful expressions}\label{appendix}

We collect here useful expressions needed to derive the main equations in the paper. Since the SV decomposition is unique, different mode contributions can be extracted and treated separately in
Einstein equations.

\subsection{Einstein Tensor}
\subsubsection{Background}
\be
a^2G^0_{\;0}=-3\HH^2+1\,,\quad
a^2G^a_{\;b}=\left(\HH^2-2\frac{a''}{a}\right)\delta^a_b\,,\quad
a^2G^z_{\;z}=\HH^2-\frac{2a''}{a}+1\,.
\ee
The time index is denoted by 0.

\subsubsection{Perturbations}
In the expressions below we keep general the value of the curvature constant $\kappa$. However, one can always replace it with $-1/2$.
\bea
\label{dg00}
a^2\delta G^0_{\;0} &=& 6\HH^2\Phi-2\HH\partial_z\Pi-2\HH \Lambda'+D^cD_c\Lambda+4\kappa\Psi-4\HH \Psi'
+2\partial^2_{z}\Psi+D^cD_c\Psi\,, \\
\label{dg0a}
a^2\delta G^0_{\;a} &=& D_bD_{[a}\bar{\Gamma}^{b]}-\frac{1}{2}\partial^2_{z}\bar{\Gamma}_a
-2\kappa\bar{\Gamma}_a+\frac{1}{2}\partial_z\partial_a\Pi-\frac{1}{2}\partial_z\bar{\Omega}'_a
+\partial_a\Lambda'\nonumber\\
&& -2\HH\partial_a\Phi+\partial_a\Psi'\,,\\
\label{dg03}
a^2\delta G^0_{\;z} &=& -2\HH\partial_z\Phi+2\partial_z\Psi'-\frac{1}{2}D^aD_a\Pi\,,\\
\label{dgab}
a^2\delta G^a_{\;b} &=&  \delta^a_b\left[2\Phi(\HH^2+2\HH')+2\HH \Phi'+\partial^2_{z}\Psi-\Psi''-2\HH \Psi'
-2\HH\partial_z\Pi -\Lambda''-2\HH \Lambda'\right. \nonumber \\
&& \left.+D^cD_c\Phi+\partial^2_{z}\Phi-\partial_z\Pi'+D^cD_c\Lambda \right]-D^{(a}D_{b)}(\Lambda+\Phi)
+D^{(a}\partial_z\bar{\Omega}_{b)} \nonumber \\
&& + D^{(a}\bar{\Gamma}'_{b)}+2\HH D^{(a}\bar{\Gamma}_{b)}\,, \\
\label{dg3a}
a^2\delta G^{z}_{\;a} &=&  -\partial_a(\partial_z\Phi+\partial_z\Psi)+\HH\partial_a\Pi
+\frac{1}{2}\partial_a\Pi'
+\HH\partial_z\bar{\Gamma}_a+\frac{1}{2}\partial_z\bar{\Gamma}_a'\nonumber\\
&& + \frac{1}{2}\bar{\Omega}_a'' +\HH\bar{\Omega}_a'-\kappa\bar{\Omega}_a
-\frac{1}{2}D_{c}D^{c}\bar{\Omega}_{a}\,,
\\
\label{dg33}
a^2\delta G^z_{\;z} &=& 2\Phi(\HH^2+2\HH')+2\HH \Phi'+D^cD_c\Phi-2(\Psi''+2\HH \Psi'-2\kappa\Psi)
+D^cD_c\Psi\,.
\eea
As usual, indices between square brackets are anti-symmetrized.

A useful expression in the derivation of equations for the vector modes is
\[
D_cD_a\bar{Z}^c=-\bar{Z}_a\,,
\]
where $\bar{Z}_a$ is any transverse vector. This identity follows from the commutator of two covariant derivatives and from the fact that $R^{(2)}_{ab}=-\gamma_{ab}$ for the pseudo-sphere.

\subsection{Energy-Momentum Tensor}

\subsubsection{Background}

The non-zero components of the anisotropic scalar field $\vartheta$ and the inflaton field are:
\be
a^2T^0_{(\vartheta)\;0}=-\frac{C^2}{2}\,,\quad
a^2T^a_{(\vartheta)\;b}=-\frac{C^2}{2}\delta^a_b\,,\quad
a^2T^3_{(\vartheta)\;z}=\frac{C^2}{2}\,.
\ee
\be
a^2T^0_{(\varphi)\;0}=-\left(\frac{\varphi'^2}{2}+a^2V\right)\,,\quad
a^2T^a_{(\varphi)\;b}=\left(\frac{\varphi'^2}{2}-a^2V\right)\delta^a_b\,,\quad
a^2T^z_{(\varphi)\;z}=\left(\frac{\varphi'^2}{2}-a^2V\right)\,.
\ee

\subsubsection{Perturbations}
The non-zero components of the perturbed energy-momentum tensor are:
\bea
a^2\delta T^0_{(\vartheta)\;0}&=&\Lambda-\partial_{z}\delta\phi\,,\\
a^2\delta T^0_{(\vartheta)\;z}&=&-\Pi-\delta\phi'\,,\\
a^2\delta T^a_{(\vartheta)\;b}&=&\left(\Lambda-\partial_{z}\delta\phi\right)\delta^a_b\,,\\
a^2\delta T^z_{(\vartheta)\;a}&=&\partial_a\delta\phi\,,\\
a^2\delta T^z_{(\vartheta)\;z}&=&-\Lambda+\partial_{z}\delta\phi\,.
\eea
and
\bea
a^2\delta T^0_{(\varphi)\;0}&=&-\varphi'\delta\varphi'+\Phi\varphi'^2-a^2V'\delta\varphi\,, \\
a^2\delta T^0_{(\varphi)\;a}&=&-\varphi'\partial_a\delta\varphi\,, \\
a^2\delta T^0_{(\varphi)\;z}&=&-\varphi'\partial_z\delta\varphi\,, \\
a^2\delta T^a_{(\varphi)\;b}&=&(\varphi'\delta\varphi'-\Phi\varphi'^2-a^2V'\delta\varphi)\delta^a_b\,, \\
a^2\delta T^z_{(\varphi)\;z}&=&\varphi'\delta\varphi'-\Phi\varphi'^2-a^2V'\delta\varphi\,.
\eea

\subsection{Klein-Gordon equations}
\subsubsection{Perturbations}
The perturbed Klein-Gordon equations for $\delta\vartheta$ and $\delta\varphi$ can be directly found
by means of $\nabla_{\mu}T^{\mu}_{\;\nu}=0$. At first order in perturbations, they are given respectively by:
\bea
\delta\vartheta''-\nabla^2\delta\vartheta+2\HH\delta\vartheta'+\Pi'+2\HH\Pi
-\partial_{z}(\Phi+2\Psi-\Lambda)&=&0\,,\\
\delta\varphi''-\nabla^2\delta\varphi+2\HH\delta\varphi'-\Phi'\varphi'-4\HH\varphi'\Phi-2\Phi\varphi''+2\Psi'\varphi'\nn \\
+\varphi'\partial_z\Pi+\varphi'\Lambda'+a^2\delta\varphi V_{,\varphi\varphi}&=&0\,,
\eea
where $V_{,\varphi\varphi}$ stands for ${\rm d}^2V/{\rm d}\varphi^2$.

\bibliography{shear-free}
\bibliographystyle{h-physrev}

\end{document}